\documentclass[12pt, preprint2]{aastex}
\usepackage[]{natbib}

\begin{document}
 
\title{New Clues to the Impact Broadening Mystery in Radio Recombination Lines}

\author{M.B. Bell\altaffilmark{1}}

\altaffiltext{1}{Herzberg Institute of Astrophysics,
National Research Council of Canada, 100 Sussex Drive, Ottawa,
ON, Canada K1A 0R6;
morley.bell@nrc-cnrc.gc.ca}

\begin{abstract}

Problems where impact broadened radio recombination lines appeared narrower than predicted first showed up $\sim40$ years ago at frequencies below $\sim3$ GHz. But it was soon found that the observations could be explained by throwing out the uniform density models and replacing them with variable density ones. However, this problem re-appeared recently when a mysterious line narrowing above quantum numbers of ($n,\Delta n$) = (202,8) was reported from sensitive observations of Orion and W51 near 6 GHz.
Here it is demonstrated that the narrowing is unlikely to be caused by the data processing technique and therefore must be source related. It is further demonstrated that the observed line narrowing can be tied to one of the fundamental properties of radio recombination lines; namely the fact that the spacing of adjacent $n$-transitions increases with frequency. The line narrowing is observed to begin when the $n$-transition density, $D_{n}$, exceeds $\sim11.6$ transitions per GHz. This may imply that it is somehow related either to a previously overlooked effect in the impact broadening process, or to some unknown parallel process, that is tied to the separation between adjacent $n$-transitions. Based on these results it can be concluded, as has also been concluded in several theoretical investigations, that the observed line narrowing is not tied to a fixed range of either $n$ or $\Delta n$. 

\end{abstract}


\section{Introduction}

Using radio recombination lines (RRLs) we can obtain much new information extending all the way from the physical conditions inside galaxies, down to the details of processes that occur inside individual atoms. As was discussed by \citet{gor08} they can thus be a valuable tool for astronomers and physicists. That it would be possible to detect RRLs was first suggested by \citet{kar59}. The theory of impact broadening of the radio lines was developed by \citet{gri67}. A broader coverage of RRLs was carried out by \citet{gor02}.  However, it soon became apparent \citep{bro72} that there were problems. \citet{hje70} had found for several nebulae that for homogeneous models, an electron density between $N_{e} \sim 2.5\times10^{4}$ and $1\times10^{5}$ is required to explain the observed recombination line intensities. But \citet{chu71} found much less broadening below $\sim 3$ Ghz than was predicted for this density and concluded that the widths predicted by Griem's theory appeared to be over-estimated at these low frequencies. It was concluded that for uniform density models the observational results for the line-to-continuum ratios were not compatible with the line profiles \citep{bro72}. However, it was argued that the failure to detect broadening at low frequencies ($\nu < 1.6$ GHz), and the smaller amount of excess broadening detected at $\nu < 5$ GHz than expected from a simple application of Griem's formula, were explicable in terms of density fluctuations in Orion \citep{chu71}, and models in which the electron density varied with radius were devised to explain the observations.

\section{More Recent Results}

\subsection{Data at 6 GHz}

The low-frequency result discussed above was not the only place where mysterious, low impact broadening results were obtained. Using a novel, multiple-overlap frequency-switching observing technique we were able to observe for the first time \citep{bel97,bel00,bel11} a series of Rydberg-Rydberg transitions in Orion and W51 near 6 GHz that ranged from ($n$,$\Delta n$) = (102,1) to (289,25) and were almost continuous in increasing $\Delta n$. When observations are carried out in a relatively narrow window and therefore close to the same frequency, the transitions available must all have different $\Delta n$ values. The ($n,\Delta n$) transitions included in our window were (102,1), (129,2), (147,3), (174,5), (184,6), (194,7), (202,8), (210,9), (217,10), (224,11), (230,12), (236,13), (241,14), (247,15), (252,16), (257,17), (261,18), (266,19), (270,20), (274,21), (278,22), (282,23), (286,24), and (289,25).

Because we used the frequency switching observing technique to obtain flat baselines without the need to fit and remove sinusoids or polynomials, the linewidths were affected by the data processing and a special procedure had to be used to recover the true linewidths \citep{bel97,bel00}. To obtain the true linewidths we first generated test Voigt profiles based on accepted impact broadening theory for each $\Delta n$-value, using the known parameters of each source, such as temperature and electron density. We then processed these test lines in a manner identical to the way the line profiles obtained at the telescope were processed. These processed test linewidths were then compared to the linewidths obtained for the processed telescope lines. As further evidence that our technique is likely to be as reliable as position-switching for strong sources like Orion it is noted here that the (He171,7) line was detected in Orion at the correct level of 3mk by \citet{bel00a} using frequency switching, but was not detected in Fig 4c of \citet{ban87} using position switching. These authors attribute this failure to large systematic effects present in the measurement of lines in this strong continuum source when position switching is used.

Observed line widths obtained for Orion and W51 at 6 GHz after processing are plotted vs $n$ in Fig 1. Data are from \citet{bel11}. The curve in Fig 1 represents the linewidths obtained after processing the test lines. Below $n$ = 200 the processed telescope lines agree with the processed test lines indicating that the linewidths calculated using the impact theory give a reasonable fit to those observed at the telescope. Because the widths of the lines are significantly reduced by our reduction process it is important to note that the actual linewidths are much larger as can be seen from the upper curve in Fig 4 of \citet{bel00}. 
	
Fig 1 thus shows that the widths of the test lines and the telescope lines increase similarly with ($n$,$\Delta n$) up to (202,8), and exactly as predicted from theory. Above (202,8) the processed test lines showed no sign of decreasing in width. However, the widths of the lines obtained at the telescope can be seen to decrease abruptly for both the Orion and W51 data. The relevant result here is the fact that the telescope profiles narrowed abruptly above $n$ = 200 while the test profiles, processed in an identical manner, did not. This appears to rule out the possibility that our reduction process was responsible for the line narrowing above $n$ = 200 since it did not affect the test lines. Further proof of this claim is presented in \S4 below.

As was found for the observations below $\sim3$ GHz, the 6 GHz data also show an unexplainable line narrowing. This mysterious result has since been examined by several authors \citep{hey06,hey07,hey09,wat06,oks04,gri05,gig07,gav07,gor08,pro10} but all attempts to explain it as an impact narrowing for quantum numbers above $n$ = 200 and above $\Delta n$ = 8 have been unsuccessful. Although the reduced impact broadening below 3 GHz might be explainable by variable density models, this is unlikely to be the explanation for the decrease seen at 6 GHz since these observations were all made with the same telescope, the same beamwidth, and near the same frequency.

\subsection{Data at 17.6 GHz}

Because of the correlation between $n$ and $\Delta n$, it is then also true that the line narrowing observed may be tied to $\Delta n$ instead of $n$ since a similar result will be obtained if the linewidths in Fig 1 are plotted versus $\Delta n$. In this case the line narrowing will begin at $\Delta n$ = 8. However, in Fig 2 the linewidths observed for Orion at 17.6 GHz are plotted versus $n$. The numbers give the $\Delta n$-value associated with each transition. Since the $n$-values do not go above 200 they give no new information. However, here the widths of the lines show no sign of narrowing even though $\Delta n$-values as high as $\Delta n$ = 17 have been detected. This means that we cannot argue that the line narrowing is tied to a given $\Delta n$-value, and this agrees with what was concluded by \citet{wat06}. Because of this, it also seems unlikely that the narrowing is tied to a particular $n$-value. What is required is a parameter that can be shown to fit all the data, explaining not only 1), the line narrowing above (202,8) at 6 GHz, but also 2), the lack of line narrowing for $\Delta n$ lines above 8 at 17.6 GHz, as well as 3), the narrowing at observing frequencies below $\sim3$ GHz.

\section{Density of Adjacent $n$-Transitions in Frequency Space}

One of the most obvious properties of RRLs is the fact that the spacing between adjacent $n$-transitions decreases significantly as the observing frequency decreases. This can be seen immediately from recombination line tables \citep{lil68,tow93,tow96}. Here we define the density of the $n$-transitions, $D_{n}$, in number of lines per GHz, as

$D_{n}$ = 1000/[$\nu_{(n,\Delta n)}$ - $\nu_{(n+1,\Delta n)}$]

where $\nu$ is the frequency of the transition in MHz.

It should be noted that $D_{n}$ is much smaller than $D_{t}$, where  $D_{t}$ is the density of all lines in a given frequency interval. Unlike $D_{n}$, $D_{t}$ does not vary with $\Delta n$ for a fixed observing window.

In Figs 3 and 4, $D_{n}$ is plotted versus $n$ and $\Delta n$ respectively for several different observing frequencies indicated by the labels. First, it is clear from these plots that the line density, $D_{n}$, varies significantly with both $n$ and $\Delta n$, as well as with the observing frequency. The horizontal line at $D_{n}$ = 11.6 in each figure represents the density at which impact broadening at 6 GHz was observed to begin to decrease and corresponds to $n$ = 202 and $\Delta n$ = 8. 

In Fig 4, the $\Delta n$ transitions detected by different investigators at several different observing frequencies have been indicated (the filled dots are the transitions detected by \citet{smi84} at 5 GHz, the open circles represent the transitions detected by \citet{bel11} at 6 GHz, the filled squares those detected by \citet{bel11} at 17.6 GHz, the open squares are detections made by \citet{roo84} at 8.7 GHz). The stars indicate the locations of the H$n\alpha$ transitions observed by early investigators. Only for line densities above $D_{n}$ = 11.6 was the impact broadening less than predicted. Recently, the GBT was used to detect Voigt profiles in several H$n\alpha$ lines in Sagittarius B2 \citep{pro10}. Since the $D_{n}$ values of these lines were all less than 10 they cannot give us any new information on line narrowing mystery.

It is clear that if $D_{n}$ = 11.6 is the density at which the impact effect begins to be reduced, the 6 GHz data would show the line narrowing while the 17.7 GHz data would not, even though lines with $\Delta n$-values up to 17 were observed. This then can explain the first two mysteries.

Finally, in Fig 3 the filled circles represent the H$n\alpha$ transitions observed at each of the different observing frequencies. The $D_{n}$-values  for these points have been re-plotted versus observing frequency in Fig 5. Here it is clear that for low-$\Delta n$ transitions, if it is true that the impact effect is reduced above $D_{n}$ = 11.6, a reduction in line width is predicted to begin below $\sim$3.5 GHz. This is exactly what was found by the early investigators and would mean that some process other than the proposed variable density model might be the real explanation for their observations.

The increasing line density model can thus be tied to the effects observed in all of the above three cases.

\section{Discussion}

Although it has been demonstrated above that the line narrowing at 6 GHz might be an effect related to the density of $n$-transitions, other possibilities need to be considered. We now consider whether a), all detections claimed are secure detections and the line narrowing is somehow related to the collision process through the $n$-transition density $D_{n}$, or whether b), all detections claimed are secure detections but the narrowing is somehow related to the data processing technique, or whether c), the high-$\Delta n$ lines that are narrower than predicted are spurious detections.

Previously we have concluded from the way the line areas in Orion and W51 are observed to fall off smoothly and as expected with increasing $\Delta n$, that the high-$\Delta n$ transitions are valid detections, which immediately rules out possibility c). The questions that remain to be answered are then: What causes the line narrowing? Does it occur in the source, or in the data reduction process? With increasing line density now seeming to play a role it is legitimate to ask whether this line narrowing could somehow be caused by the reduction process, possibly related to line confusion, and we now consider this possibility.

First, as noted above, the total line density $D_{t}$ for all $\Delta n$ values must be much higher than the density of the $n$-transitions related to one $\Delta n$ value. At a given frequency the total line density for all $\Delta n$ lines between 1 and 20 is easily obtained by counting the lines listed in recombination line tables. For 6 GHz this value is $\sim263$ lines per GHz, which corresponds to an average of 1 line every 3.8 MHz. For linewidths close to 0.5 MHz this would not normally lead to confusion. However, if $\Delta n$ values greater than 20 are included, and the observed lines are immersed in a low-level background of faint lines, confusion could be a problem. A simple analysis shows that most of the noise from this background is at low spatial frequencies, unlike continuum noise, which could selectively bury the low spectral frequencies from the Voigt profile wings. The detectable signal would thus be dominated by high spatial frequencies, which are primarily in the central, narrow part of the lines, resulting in only the "tip of the iceberg" as it were, being detected.

Although $D_{t}$ is constant and does not vary with $\Delta n$, the lines do get weaker with increasing $\Delta n$ and the wings get wider. If this is the correct interpretation it would lead to a significant loss of power in the lines as $\Delta n$ increases, as well as the observed decrease in the linewidth. Furthermore, this effect would be present only for the processed telescope lines and absent for the processed test lines where there are no confusing lines, so the line narrowing would also be expected to show up only in the telescope lines as seen in Fig 1.

Whether or not this is the correct interpretation can easily be checked by comparing how the power in the line changes with increasing $\Delta n$ for both the telescope lines and the test lines. If the line narrowing with increasing $\Delta n$ is due to this confusion-related process and only the upper tip of the line is being detected, the power detected in the lines must then decrease faster for the telescope lines than for the test lines.

We have previously plotted the line area versus $n$ on logarithmic scales for both the Orion and W51 data and found that \em the opposite is true\em. The area of the telescope lines falls off more slowly than the area of the test lines \citep[see their Figs 3 and 5]{bel00}. This indicates that the line confusion process cannot be the cause of the line narrowing, leaving us with the likelihood then that the increasing line density model discussed above is the true explanation for the line narrowing.

In our previous paper \citep{bel00} we expressed concern over why the processed telescope line areas remained above the processed test line areas while the line widths narrowed for $\Delta n > 8$ at 6 GHz. There is now a simple explanation for this mystery. The dashed line in Figs 3 and 5 of \citet{bel00} show how the true areas of the lines are expected to decrease with increasing $n$ and $\Delta n$. The solid curve in Figs 3 and 5 shows how the processed line area is reduced for the test lines by the overlapping process associated with frequency switching. This reduction increases with $\Delta n$ \em because the linewidth increases. \em If something causes the line narrowing to occur before the data reduction takes place, but without loss of power in the lines, since narrower lines are less affected by the overlapping process the line areas would then be predicted to fall off \em slower \em than the test lines, as found. This slower fall-off is then exactly the evidence required to confirm that the linewidth is truly increasing less rapidly with $\Delta n$ in the source lines than in the test lines, which indicates that the line narrowing process must be source related, taking place \em prior to the data reduction process. \em

It is concluded that all the evidence then indicates that the line narrowing seen in impact broadened recombination lines is source related occurring prior to the data reduction stage. The evidence also shows that it can be linked directly to the increase in the $n$-transition density $D_{n}$ that occurs as the observing frequency decreases.

Because this effect is tied to the density of adjacent $n$-transitions, as opposed to the total line density in frequency space, this may suggest that the effect is related to the fact that in a collision it is only those $n$-levels that are close to one another that will be affected while more distant $n$-levels are not involved, even though they may be part of transitions that produce lines in the same frequency space.

Finally, we might also ask if the effect we are discussing could be due to masering. This would appear to be unlikely. The level populations and their inversions at high $n$ are well studied and appear to rule out strong maser amplification at cm wavelengths \citep{str96}. Strong line masering should occur only at mm wavelengths in regions of large mm optical depth where the combined line and continuum optical depth can be large and negative. These conditions are rare and are convincingly seen in only one or two sources (e.g. MWC 349). The lines are also variable. At cm wavelengths, the net optical depths (line + continuum) are always positive, preventing strong maser lines. There can be a modest enhancement of the line intensity which is produced by a reduced absorption of the associated continuum at the line centre, and which then appears as a slight enhancement in net emission at the line centre over that of LTE. The enhancements are usually $10-30\%$ but there is no line narrowing.

\section{Conclusion}

The instances where impact broadened linewidths have been found to be narrower than expected have been re-examined. It is demonstrated that the line narrowing for high $\Delta n$ values at 6 GHz must have occurred prior to the data processing stage and therefore must be source related. It is also demonstrated that it is unlikely to be due to line confusion. It is further demonstrated that the observed line narrowing can be tied to one of the fundamental properties of radio recombination lines; namely the fact that the spacing of adjacent $n$-transitions increases with frequency. The line narrowing begins when the $n$-transition density, $D_{n}$, exceeds $\sim11.6$ transitions per GHz. This may imply that it is somehow related either to a previously overlooked effect in the impact broadening process, or to some unknown parallel process, that is also tied to the $n$-transition line density. Based on these results it can also be concluded, in agreement with several previous theoretical investigations, that the line-narrowing effects are not tied to any particular range of either $n$ or $\Delta n$.

\section{Acknowledgements}

I thank Ernest Seaquist and James K.G. Watson for many helpful comments.

\begin{figure}
\hspace{-1.0cm}
\vspace{-1.0cm}
\epsscale{0.9}
\includegraphics[width=9cm]{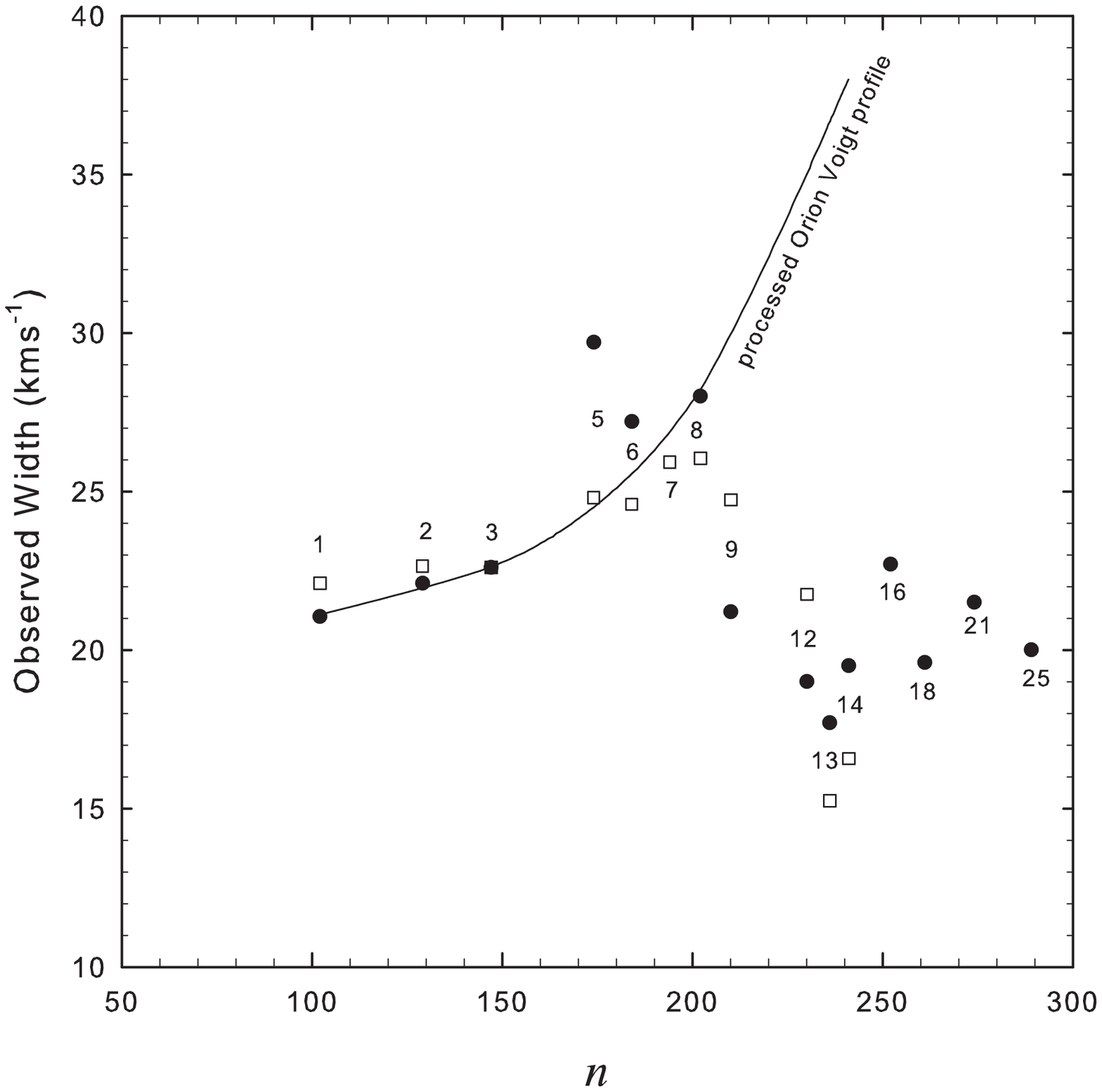}

\caption{{Observed line width at 6 GHz from \citet{bel11} after processing, plotted vs $n$ for Orion (filled circles) and W51 (open squares). The numbers represent the relevant $\Delta n$ values. The curve represents the line widths obtained using the Voigt profile test lines generated using impact broadening theory and known source parameters. There is no evidence that the reduction process has affected the width of the test lines above $n$ = 200. Here $n$ is the value of the lower level of the transition. $\Delta n$ = 8 corresponds to $n$ = 202 in this figure because ($n,\Delta n$) = (202,8) is the $\Delta n$ = 8 transition whose frequency falls inside our observing window. \label{fig1}}}
\end{figure}

\begin{figure}
\hspace{-1.0cm}
\vspace{-1.0cm}
\epsscale{0.9}
\includegraphics[width=9cm]{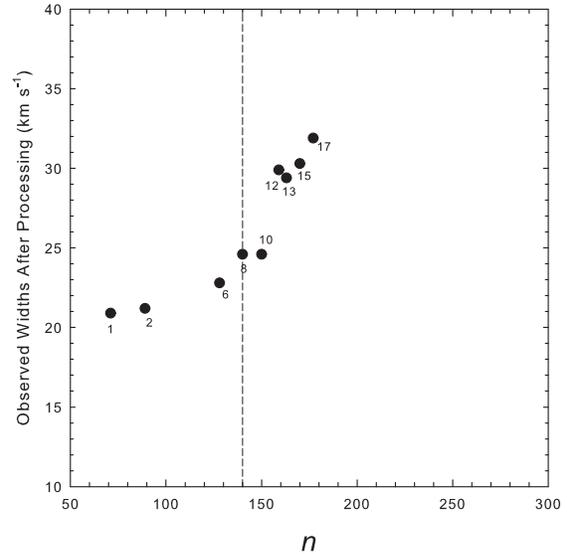}
\caption{{Observed line width at 17.6 GHz from \citet{bel11} after processing, plotted vs $n$ for Orion. Numbers are the associated $\Delta$$n$-values. They show no sign of narrowing above $\Delta n$ = 8 which is indicated by the dashed line. \label{fig2}}}
\end{figure}

\begin{figure}

\hspace{-1.0cm}
\vspace{-1.0cm}
\epsscale{0.9}
\includegraphics[width=9cm]{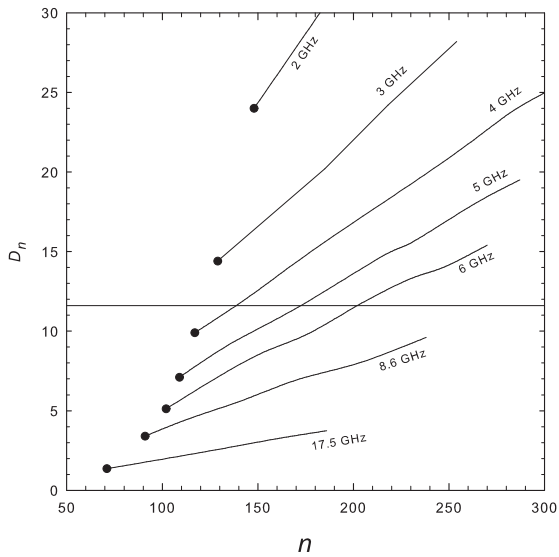}
\caption{{Plot showing how $D_{n}$ changes with $n$ for different observing frequencies. Impact broadening is reduced above a line density of 11.6 lines per GHz which occurs at $n$ = 200 for lines near 6 GHz. Thus no narrowing would be expected at 17 GHz, while little broadening would be expected at 1.4 GHz. The filled dots represent H$n\alpha$ lines at each frequency. \label{fig3}}}
\end{figure} 
 
\begin{figure}
\hspace{-1.0cm}
\vspace{-1.0cm}
\epsscale{0.9}
\includegraphics[width=9cm]{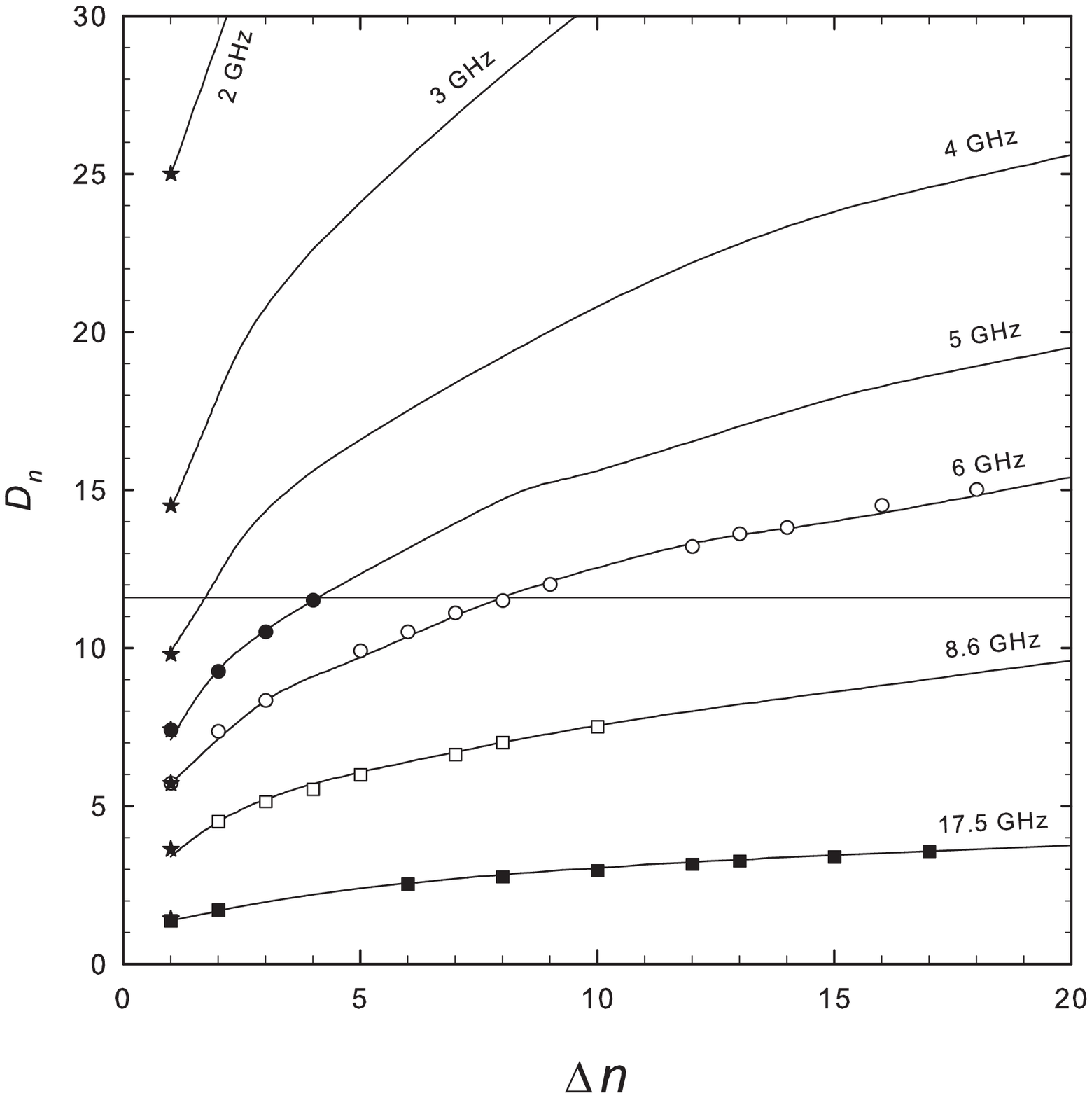}
\caption{{Plot showing how $D_{n}$ changes with $\Delta$$n$. (filled circles) the transitions detected at 5 GHz from \citet{smi84}. (open circles) the 6 GHz transitions from \citet{bel11}. (filled squares) the 17.6 GHz transitions from \citet{bel11}. (open squares) the 8.7 GHz transitions detected by \citet{roo84}.(stars) H$n\alpha$ lines. If impact broadening begins to be reduced above $D_{n}$ = 11.6 it would only be apparent so far in the 6 GHz data and the low frequency H$n\alpha$ lines. \label{fig4}}}
\end{figure}
 
\begin{figure}
\hspace{-1.0cm}
\vspace{-0.5cm}
\epsscale{0.9}
\includegraphics[width=9cm]{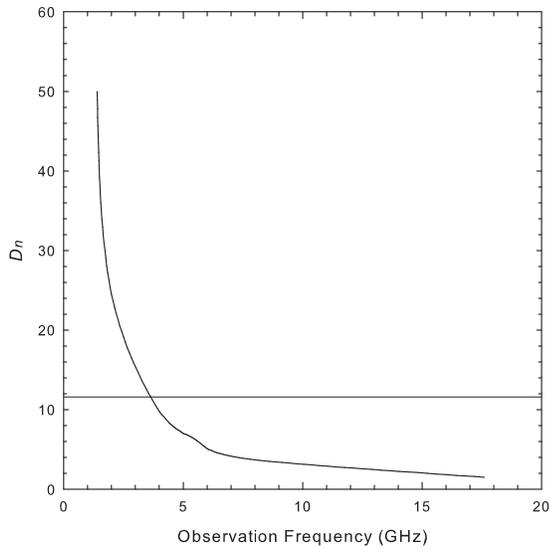}

\caption{{Plot showing how the line density for alpha lines changes with observing frequency. Impact broadening is reduced above $D_{n}$ = 11.6 lines per GHz which occurs for Hydrogen alpha lines observed below $\sim3.5$ GHz, as found by early investigators. \label{fig5}}}
\end{figure}

\end{document}